\documentclass[dvips]{astavr}
\usepackage{supertabular,lscape,epsfig}
\usepackage{amssymb}
\usepackage{amsmath}

\SetPages{1}{1}
\SetVol{65}{2015}

\begin{document}
\begin{Titlepage}

\Title{ On the global and local magnetic fields in
flare stars. Study of YZ CMi and OT~Ser }

\Author{V.D. Bychkov$^{1}$, L.V. Bychkova$^{1}$, J. Madej$^{2}$, A.A. Panferov$^{3}$ }
{$^{1}$ Special Astrophysical Observatory, Russian Academy of Sciences  \\   
       Nizhnij Arkhyz, 369167 Russia    \\
$^{2}$ Warsaw University Observatory, Al. Ujazdowskie 4, 00-478 Warszawa, Poland   \\
$^{3}$ Togliatti State University, Belorusskaya 14, Tolyatti, 445667 Russia }

\end{Titlepage}

\Abstract{
Global magnetic fields of flare stars can evolve rapidly, in time
scale of hundreds or dozens of days. We believe, that such changes 
result from rapid superposition of local magnetic fields generated 
by differential rotation of those stars. We discuss possible mechanisms
of generation and dissipation of local and global magnetic fields in
sample flare stars OT Ser and YZ CMi. Mechanism of magnetic
braking of these stars is proposed here, in which differential rotation generates local
magnetic fields, and eventually energy accumulated in local fields is
radiated away by flares. We obtained estimates of the rotational energy
and the energy of the global magnetic field of OT Ser and YZ CMi. 
It is shown that the energy of the local magnetic fields dissipated
during superflare of YZ CMi on 9 February 2008 (UT 20:22:00) did not
influence the global magnetic field of this star. 
}

{Stars: chemically peculiar -- Stars: magnetic fields }

  \begin{flushright}
    \vspace{0cm}
    {\it submitted to } {\rm Astrophysical Bulletin } \\
  \end{flushright} 

\section{Introduction}

Approximately 70 \% of stars in our Galaxy are red dwarfs. Majority of
stars in that group exhibit flare activity. Such a bright and intriguing
form of light variations has drawn attention of many reserchers.
Consequently, there are published many papers on that subject and thousands
of red dwarfs were investigated up to now. For few objects the phenomena
of flare light variations were observed in the full range of wavelengths,
see Gershberg et al. (1999, 2011) and Katsova et al. (1999).

Our understanding of the nature of such light variations of red dwarfs 
was extended by the analogy to similar events observed on the Sun. Models
comprehensively describing eruptions were presented by Hawley et al. (1995),
Katsova et al. (1999), Katsova \& Livshits (2001), Shibata \& Yokoyama 
(1999; 2002), Stepanov et al. (2005) and others. However, there exist new
observational data which cannot be explained by the existing models.

Flare stars show an incessant generation of local magnetic fields on the
surface. In atmospheres of such objects energy of rotation and energy of
convective motion of matter partly transforms into energy of the magnetic
field. It is a reasonable assumption, that the power of such a magnetic
field generator is constant at a given evolutionary stage and can change only
when a star change due to its evolution. The above prediction was frequently
formulated by R.E. Gershberg. Average power of the generator is determined
by the rotational velocity of a star, the measure of differential rotation,
the effective temperature and mass of that star, Rossby number etc.

Other important measurable quantity is the total energy released during a
flare. Accuracy of such a time-integrated measurement strongly depends on
the time distribution of individual observations. Moreover, some flares
remain undetected at all. Some flares occur at the rear side of a star.
Some of them can be only partly visible to an observer and therefore can 
change their apparent properties due to the obscuration by a star.

At present there exist various techniques to account for and correct the 
above uncertainties, and we can infer the averaged power of the magnetic
field generator. That quantity seems well estimated for the most observed
M stars using, for instance, statistical investigations by Gershberg (1972), 
Moffett (1974), Lacy et al. (1976) and Kowalski et al. (2010). 

Lacy et al. (1976) obtained the following relation for the best investigated 
flare stars
\begin{equation}
\log \nu = \alpha + \beta \, E_U \, ,
\end{equation}
where $\nu$ denotes frequency of eruptions in hours$^{-1}$, variable $E_U$ is
the energy of eruption as seen in the U filter of the Johnson wideband UBV 
photometry. Note here, that the dominant fraction of flare energy is released
in short wavelengths. Assuming estimated values of $\alpha$ and $\beta$ from
Lacy et al. (1976) one can estimate in a rough approximation the lower limit
for the averaged power of the generator of local magnetic fields for selected
flaring stars. I.e. this is the estimate of energy emitted per 1 second as
seen in the U filter.

\begin{table}[h]
\label{table:1}
\caption{ Estimates of the power of the local magnetic fields generator
   for selected flare stars. }
\vspace{2mm}
\begin{tabular}{|l|c|c|c|}
\hline
Star        &   $\alpha$   &  $\beta$       &  $\log~ E_{U}$ (erg/s)  \\
\hline
CN Leo      & 28.6  $\pm$ 3  & -0.99 $\pm$ 0.12 & 25.297 $\pm$ ~ 4.375 \\
UV Cet      & 29.0  $\pm$ 4  & -0.98 $\pm$ 0.14 & 25.963 $\pm$ ~ 5.624 \\
Wolf 424 AB & 24.1  $\pm$ 5  & -0.81 $\pm$ 0.18 & 25.363 $\pm$ ~ 8.784 \\
YZ CMi      & 21.4  $\pm$ 2  & -0.71 $\pm$ 0.08 & 25.132 $\pm$ ~ 4.046 \\
EQ Peg A    & 30.7  $\pm$ 4  & -1.00 $\pm$ 0.14 & 27.144 $\pm$ ~ 5.626 \\
EV Lac      & 20.7  $\pm$ 3  & -0.69 $\pm$ 0.11 & 24.846 $\pm$ ~ 6.028 \\
AD Leo      & 24.4  $\pm$ 8  & -0.82 $\pm$ 0.27 & 25.419 $\pm$ 14.375 \\
YY Gem      & 12.6  $\pm$ 3  & -0.43 $\pm$ 0.11 & 21.032 $\pm$ ~ 9.350 \\
\hline
\end{tabular}
\end{table}

\section{Global magnetic fields in red dwarfs }

Global and local magnetic fields in 11 red dwarfs were first studied in detail
by Donati et al. (2008) and Morin et al. (2008; 2010). They showed, that red
dwarfs exhibit two essentially different types of magnetic field structure:

\begin{enumerate}
\item SD (strong dipolar)  -- strong and stable two-polar magnetic field,
\item WM (weak multipolar) -- weak multipole field.
\end{enumerate}
The above two types of magnetic field configurations exist in various stars
of the same masses and similar periods of rotation, see Morin et al. (2011a;
2011b).

Donati et al. (2008) and Morin et al. (2010) also discovered a puzzling 
peculiarity in magnetic behaviour of those stars: longitudinal magnetic 
field $B_l$ of some objects (quantity integrated over the visible disc 
of a star) can rise or decrease steeply. We can demonstrate this amazing
effect in case of flare star OT Ser using measurements by Donati et al. (2008). 

\begin{figure}[ht!]
\includegraphics[angle=0, width=0.8\columnwidth]{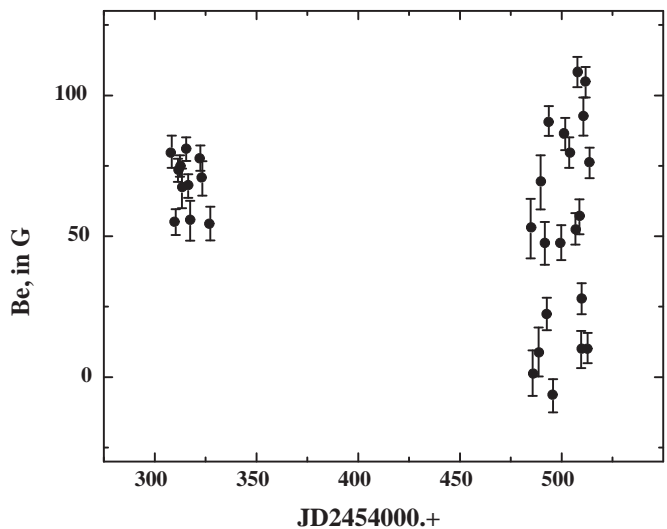}
\caption{Flare star OT Ser. Magnetic field $B_l$ as a function of $HJD$.}
\label{fig:1}
\end{figure}

{
\begin{figure}[ht!]
\includegraphics[angle=0, width=0.8\columnwidth]{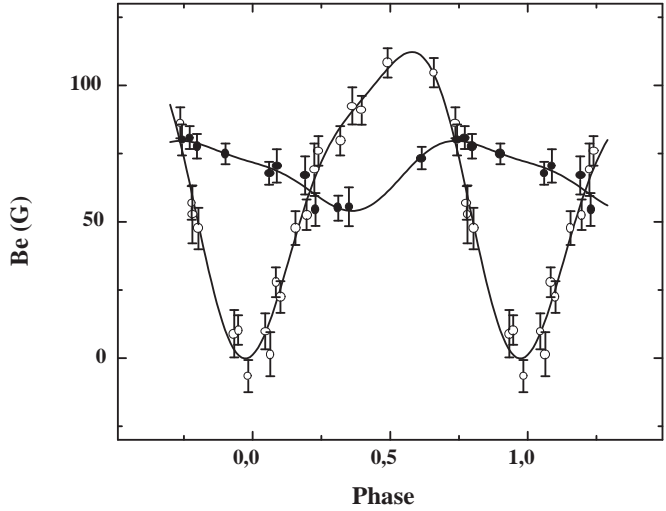}
\caption{Magnetic phase curve $B_l$ plotted against rotational phase for flare
   star OT Ser with the rotational period $3^{d}.424$. First and second
   set of observations (filled circles and open circles, respectively) 
   are separated by about 180 days. }
\label{fig:2}
\end{figure}

\begin{figure}[h]
\includegraphics[angle=0, width=0.8\columnwidth]{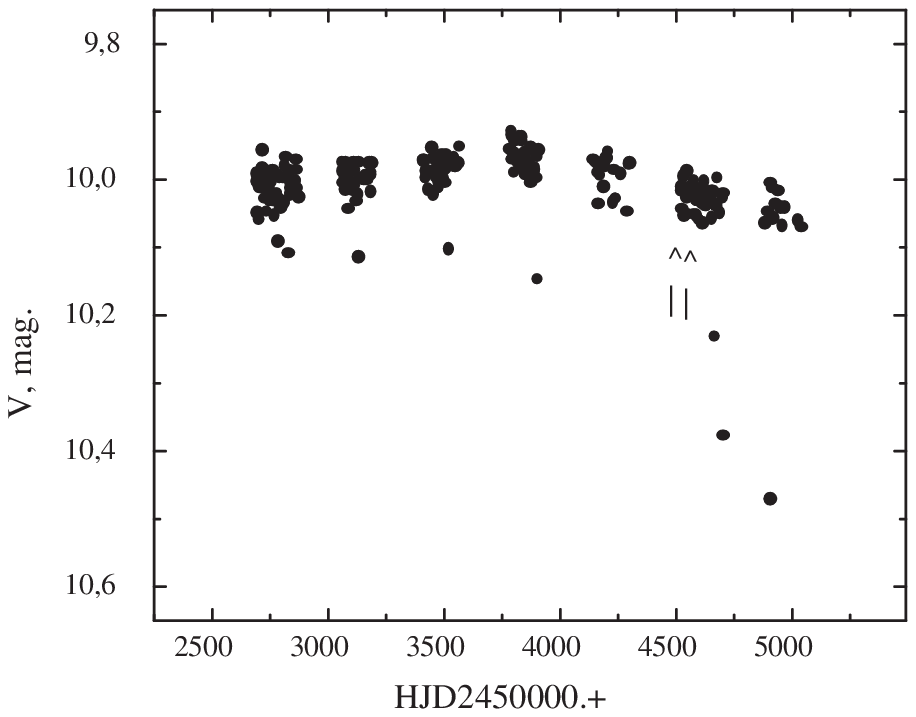}
\caption{Light variations in V color of flare star OT Ser in years 2003--2010
determined by the photometric survey ASAS3. Average times for both $B_l$
sets of the magnetic measurements are indicated by vertical arrows. }
\label{fig:3}
\end{figure}
}

Fig. 1 shows run of discrete values of the longitudinal magnetic field 
$B_l$ vs. time of observation in $HJD$. One can see that OT Ser was
accidentally observed in two different states. In the first state the
longitudinal component of the magnetic field $B_l$ integrated over stellar
disk shows low periodic variability of complex shape with amplitude 12 G
about the average value 68 G. In the
second set of measurements, which was obtained a half of the year later,
that picture dramatically changed. Values of $B_l$ rose significantly 
stepwise and now they show periodic variations with the period of rotation
equal 3.424 days and the amplitude 54 G (4.5 times higher). 

The most optimistic theoretical estimates predict the rate of evolution of
the complex global magnetic field structure (like that observed in flare
stars) of the order $10^{6} -10^{7}$ years [22]. Therefore, changes of
amplitude and shape of the magnetic phase curve within six months are
spasmodic and such rapid changes indicate that in flare stars there exists
fundamentally different mechanism of generation and evolution of the global
magnetic fields than in stars on the upper part of the main star sequence.

Magnetic phase curve for OT Ser in the second state can be approximated 
by a double sine wave
\begin{equation}
B_{e} (\phi) = B_0 + B_1 \cos (\phi+z_1) + B_2 \cos ({2\phi}+z_2) \, ,
\end{equation}
where
\begin{equation}
\phi = 2\pi \, \left( {{t_i - T_0} \over P} \right) \, .
\end{equation}
with parameters $B_0=64$ G, $B_1 =53$ G and $B_2=13$ G. 

Therefore, OT Ser has changed the type of its magnetism from WM type (weak
multipolar) to SD type (strong dipolar) during six months which separate
both series of observations. It is of great
importance to understand what happened to OT Ser in the time period between
both states and what is the reason of such a qualitative change of its
observable magnetic behaviour.

Fig. 3 shows variation of the apparent brightness of OT Ser in V color 
in years 2003--2010. Points V were taken from the database of the robotic 
photometric survey ASAS3 (Pojma\'nski 1997). OT Ser shows a smooth 
long-term vaviability of brightness in that color without any distinct
features in the light curve. Average times of both $B_l$ sets of the 
magnetic measurements of OT Ser (as in Fig. 1) are indicated here by
arrows.

\section{Local and global magnetic fields: case of YZ CMi }

Magnetic type of red dwarf YZ CMi points to SD star (strong dipolar). 
Morin et al. (2008) published measurements of this star obtained in years
2007-2008. Fig. 4 and Table 2 show, that the amplitude of magnetic variations
of this star slightly decreased at that time period.

\begin{figure}[ht!]
\includegraphics[angle=0, width=0.8\columnwidth]{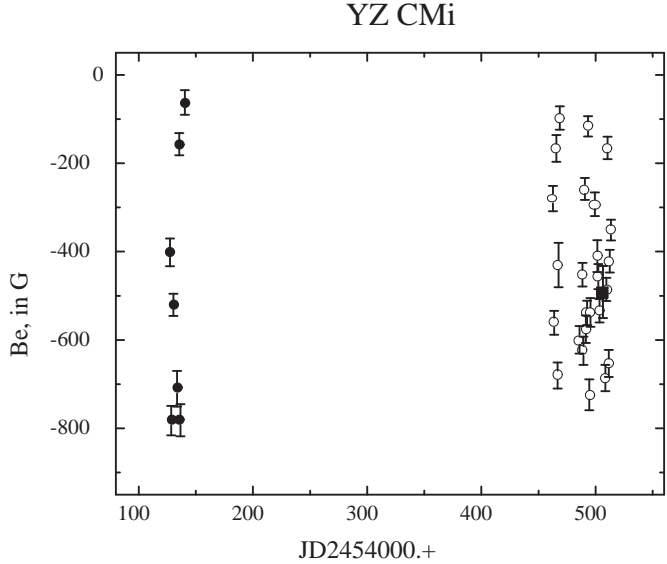}
\caption{Measurements of the longitudinal magnetic field of YZ CMi in
years 2007-2008 published in Morin et al. (2008). The second set of
magnetic measurements (open circles) presents a single $B_l$ point
(filled square) obtained just afters the end of superflare on 9 February
2008 UT 23:06:31. }
\label{fig:4}
\end{figure}

\begin{figure}[ht!]
\includegraphics[angle=0, width=0.8\columnwidth]{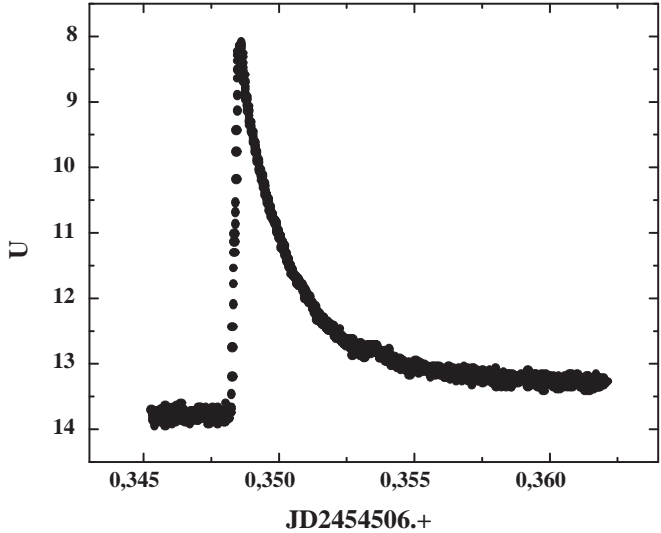}
\caption{ Apparent brightness of YZ CMi in filter U as a function of HJD
during superflare on February 9, 2008 UT 20:22:00. }
\label{fig:5}
\end{figure}

\begin{figure}[ht!]
\includegraphics[angle=0, width=0.8\columnwidth]{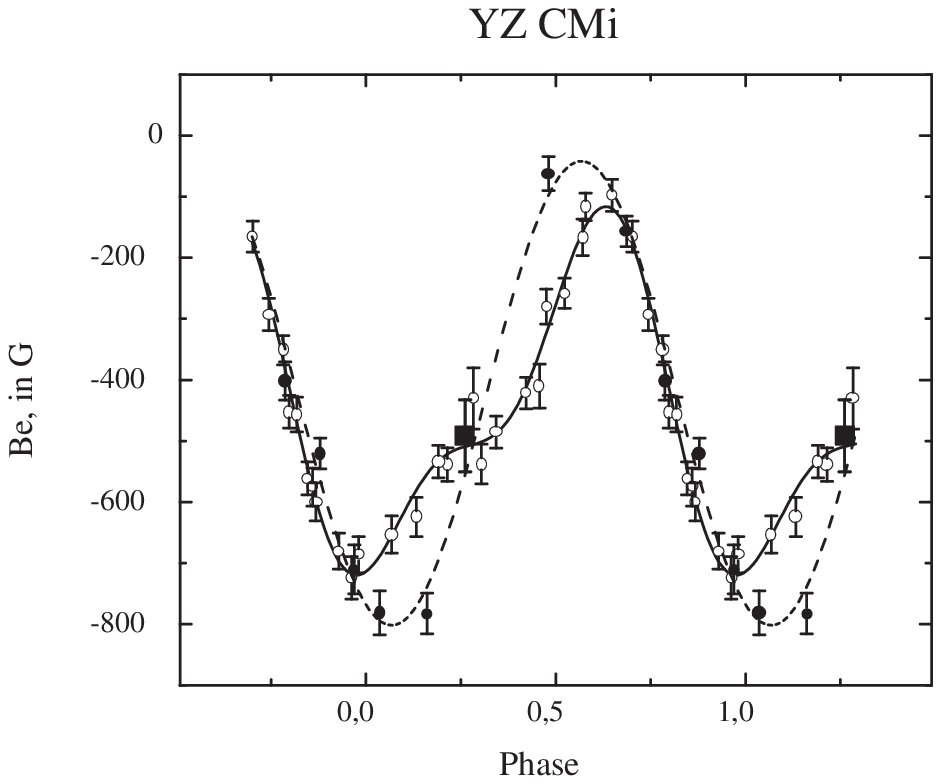}
\caption{Magnetic phase curves for flare star YZ CMi with the rotational
period $2^{d}.7773$. All $B_l$ measurements were taken from Morin et 
al. (2008). Filled circles and dashed line present data of 2007 year and
define a simple sine wave phase curve. Measurements obtained in year 2008
(open circles) define more complex double wave phase curve. Filled
square indicates a single $B_e$ observation obtained just after the end
of the superflare on 9 February 2008, UT 23:06:31. }
\label{fig:6}
\end{figure}

Fig. 6 presents magnetic phase curves for YZ CMi obtained in those years.
As is easily seen, in 2007 the phase curve was approximated by a simple
harmonic wave with the parameters $B_0 = -422\pm 14$ G, $B_1 = 300\pm 16$
G and the period $P =~2.7773$ days. Unfortunately, the total number of
$B_e$ points obtained in 2007 was not high. Phase curve obtained in 2008
shows more complex double wave structure with lower amplitudes. New 
parameters were given by $B_0 = -446\pm 6$ G, $B_1 = 247\pm 9$ G and 
$B_2 = 106\pm 9$ G.

Methods for obtaining the above phase curves are the same as in the 
catalog of the average magnetic phase curves (Bychkov et al. 2005).

On 9 February 2008 UT 20:22:00 a very strong burst started on YZ CMi. The
superflare lasted for about 1 hour and was observed with the high-speed
UBVRI photometer and the 2-m telescope at the Terskol peak observing
station (Zhilyaev et al. 2011). Then, authors estimated parameters of the
flare, and the size and other parameters of the area on the star where
that eruption occured. 

That event was among the most powerful eruptions ever observed in YZ CMi,
taking into account the energy yield. At the peak of the flare luminosity
of the star in U filter increased 180 times ! Then, the power received in
U filter reached 20\% of the bolometric luminosity of the star ($M_{\rm 
bol} =10.25$, Reid \& Hawley 2005). 

Fig. 5 presents estimates of YZ CMi brightness in the U filter during the
superflare. End of flash has not been observed and it was necessary to
extrapolate brightness variations in U filter to estimate brightness in
the quiescent state. But it could not significantly affect accuracy of 
the total yield estimate of the flare and the time length of this eruption.

Accidentally, at the end of that superflare on 9 Feb 2008 UT 23:06:31,
Morin et al. (2008) measured longitudinal magnetic field $B_l$ of YZ CMi
using the 2-m Telescope Bernard Lyot (TBL, southern France) equipped
with NARVAL spectropolarimeter. This single estimate was obtained in
the second set of $ B_{l}$ measurements for YZ CMi. In Fig. 4 this estimate
is represened by a filled square.
As can be seen in Fig. 4, even such a powerful eruption, or dissolution of the local 
magnetic field did not influence the global field structure of the star.
 
All estimates $ B_{l} $ from the second set of measurements (year 2008)
are well described by the magnetic phase curve defined by Eq. (2) with
parameters - $ B_0$, $B_1$ and $B_2 $ quoted above. In order to show 
goodness of this curve (solid line in Fig. 6) we computed corresponding
deviations of $B_l$ points from the smooth phase curve.

Fig. 6 presents deviations of the discrete longitudinal field values $B_l$
from the averaged phase curve. Position of the $B_l$
measurement obtained on 9 Feb 2008 was indicated by a solid square, same as
in Fig. 4. It is evident, that the $B_e$ point does not distinguish from
other measurements  and the apparent trends in run of other $B_l$ points
represent just random inaccuracies in fitting by a double wave phase curve.

\begin{figure}[ht!]
\includegraphics[angle=0, width=0.8\columnwidth]{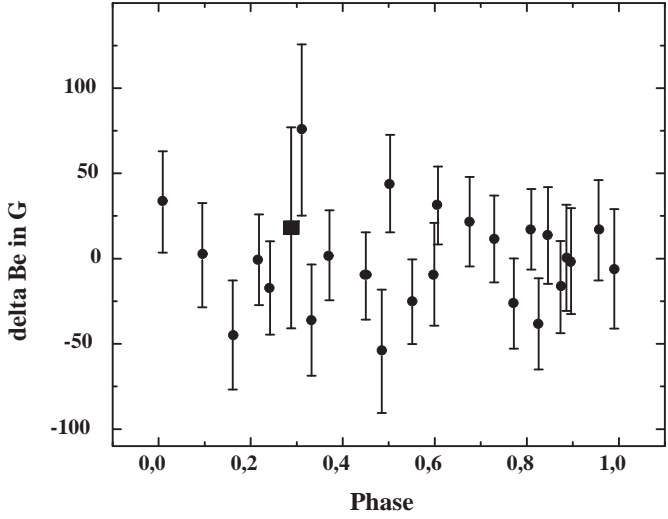}
\caption{Deviations of $B_e$ points from the phase curve for flare star 
YZ CMi with the rotational period 2.7773 days. 
The single $B_e$ point annotated by a solid square was observed just after 
the superflare has ended on 9 Feb 2008 UT 23:06:31. }
\label{fig:7}
\end{figure}

Fig. 7 shows those deviations from the magnetic phase curve (solid line).
Again, the single $ B_{l}$ point measured immediately on the end of superflare
(9 February 2008) is annotated by a large filled square. Fig. 7 clearly shows
that this single $B_l$ measurement is an average point and all deviations are
random. We conclude that even such a powerful local magnetic field dissipation
event had no effect on the global magnetic field.

Other authors also observed megabursts in this frequently observed flare
star. For example, on 16 January 2009 UT max 04:32:00 at the peak flux of
the flare luminosity of YZ CMi in U filter increased by 5.8 magnitudes (i.e.
$\sim 330$ times !), see Kowalski et al. (2010). Parameters of superflares 
differ from average flares not only by the amount of energy. Superflares
also are events of significantly longer duration (Kowalski et al. 2010).

\section{Discussion of results }

We can present approximate estimate of energy yield during the superflare
of 9 Feb 2008 (Zhilyaev et al. 2011). Following Gershberg (1972), one should
integrate the light curve over the time. That time equals approx.
$1.41 \times 10^4$ sec, i.e. it is the time period when YZ CMi radiated 
away that amount of energy in quiescent state. According to Moffett (1974)
the average flux in U filter from YZ CMi in quiescence equals
$4.00 \times 10^{28}$ erg/s. Consequently, the integrated additional energy
released at the time of the flare equals $5.64 \times 10^{32}$ erg/s. Time
necessary to accumulate such amount of energy equals ca. 1.3 years, if we
accepted the average power of the local field generator with parameters
quoted in Table 1. Such an estimate obviously contradits observations.

If we accept the upper limit for an estimate for the power of the generator,
$\log E_U \approx 29.2$ erg/s, then the time period decreases to ca. 1 hour
which is much closer to observational data. On the other hand, the observed
power of the generator was significantly overestimated here. 

We present below sample estimates of the mechanical energy of a star the
energy of its global magnetic field. Energy of the rotational motion of a
solid body equals
\begin{equation}
W_{r} = {I\omega^{2}\over 2} \, ,
\end{equation}
where $I=kMR^2$ is the moment of inertia, coefficient $k$ was determined by
the mass distribution inside a star, and $\omega=2\pi/P$ stands for angular
velocity of rotation. Exact value of the coefficient $k$ was computed for a
Lane-Emden model of a star in hydrostatic equilibrium, for politrope index
$n=3/2$ corresponding to a convective star.

Energies of the global magnetic fields of OT Ser and YZ CMi are estimated
under assumption, that the field configuration in both stars is close to a 
simple dipole. This assumption follows the observational fact, that both
magnetic phase curves are close to simple sine waves. Then, in a rough
approximation the energy of magnetic field was estimated by
\begin{equation}
W_m \approx \frac{B^2}{8\pi} \times \frac{4 \pi R^3}{3} \, \rm erg , 
\end{equation}
aditionally assuming that the field is homogeneous in the volume of a star.

Estimates of the principal parameters in Table 2, as angles $\beta$, $i$
and the polar intensity of the magnetic field $B_p$ were obtained following
the Stibbs-Preston formalism (Stibbs 1950; Preston 1971).

\begin{table}[h]
\label{table:2}
\begin{center}
\caption{ Principal parameters of flare stars OT Ser and YZ CMi. }
\begin{tabular}{|l|c|c|c|c|c|c|}
\hline
parameter            &  OT Ser              &                      &                    & YZ CMi              &                     &        \\
                     &                      &     2007             &     2008           &                     & 2007                & 2008   \\
\hline
$M/M_{\odot}$        &  0.55                &                      &                    & 0.31                &                     &        \\
Sp.type              &  M1.5~V              &                      &                    & M4.5~V              &                     &        \\
$R/R_{\odot}$        &  0.49                &                      &                    & 0.29                &                     &        \\
$k$                  &  0.00519             &                      &                    & 0.00191             &                     &        \\
$P_{\rm rot}~$in d   &  3.424               &                      &                    & 2.77729             &                     &        \\
$V \sin i$ in km/s   &  6 $\pm$ 1           &                      &                    & 5 $\pm$1            &                     &        \\
$i$ in degr.         &  56 $\pm$ 20         &                      &                    & 71 $\pm$ 20         &                     &        \\
$B_0$ in G           &                      &  68 $\pm$  2         &  65 $\pm$  2       &                     & -422 $\pm$ 14       & -453 $\pm$ 6        \\
$B_1$ in G           &                      &  12 $\pm$  2         &  54 $\pm$  3       &                     &  380 $\pm$ 16       &  250 $\pm$ 9        \\
$\beta$              &                      &     7                &  29                & 12                  &                     &                     \\
$B_{p}$ in  G        &                      &    406               &  436               &                     &  4498               &  4696               \\
$W_{r}$ (erg)        &  1.5 $\times 10^{42}$&                      &                    & 1.6 $\times 10^{41}$&                     &                     \\
$W_{m}$ (erg)        &                      &6.9$\times 10^{36}$   &1.3$\times 10^{36}$ &                     & 2.8 $\times 10^{37}$& 3.0 $\times 10^{37}$\\
$W_{m}/{W_{r}}$      &                      &4.6$\times 10^{-6}$   &0.87$\times 10^{-6}$&                     &1.75 $\times 10^{-4}$&1.88 $\times 10^{-4}$\\
\hline
\end{tabular}
\end{center}
\end{table}

Unfortunately, OT Ser was not previously included into the table of most
frequently observed flare stars and, therefore, we could not collect
enough data to estimate the power of the local magnetic field generator in
this star (see Table 1). The only solution of this problem in OT Ser is to
apply here that power for YY Gem, since both flare stars show most similar
parameters.

If one assumes, that the global magmnetic field of OT Ser was cumulated 
from the generator of local magnetic fields in the time period $\approx$
100 days, then its power should be higher than $1.5\times 10^{29}$ erg/s.

\section{Conclusions}

Estimates collected in Table 2 show, that the rotational energy of a star
is much higher than energy of its global magnetic field and certainly can
be a source of 'fuel' for the dynamo mechanism for the magnetic field 
generation. In late-type dwarfs with differential
rotation exists an efficient dynamo-like mechanism generating local
magnetic fields powered by rotational energy. Energy accumulated in
local fields is radiated into space by coronal flares. Therefore, it
is most important to identify mechanism generating differental rotation
of late type stars, which eventually continuously generates local 
magnetic fields and flares in which magnetic field energy is radiated
into space. In fact it is a consistent transformation of mechanical 
energy into magnetic energy and eventually transformation into heat.

In our opinion in late type stars simultaneously exist two essentially 
independent types of magnetic field: strong dipolar SD and weak 
multipolar WM fields. One can assume, that in some cases generated 
local magnetic fields can shape without 
dissipation and eventually can create global magnetic fields
($\alpha^2$ mechanism). The example: OT Ser which showed transition
from WM to SD state. We believe, that the opposite scenario also is
possible. In the latter case set of local fields compensates already
existing global magnetic field and weaken it. Probably it can be
seen in YZ CMi, where we noted decrease of the global magnetic field.

We expect, that the forthcoming influx of new observational data will
verify the above considerations. On the other hand, existing classical
models of flare eruptions must also be reviewed. 

New flare models certainly should take into account quite strong global 
magnetic field penetrating hot coronal gas which surrounds the flare 
region. Modeling of coronal magnetic arches and loops describing local
flares must include presence of the global field, which was found in
some red dwarfs, see Donati et al. (2008) and Morin et al. (2008; 2010).

\section{Acknowledgements}

Authors thank B.E. Zhilyaeva for providing original photometric observational data,
as well as R.E. Gershberg and I.Yu. Alexeyev for useful discussions.
This work was supported by the Russian Science Foundation
(project No. 14-50-00043).


\begin{references}

\refitem{
Bychkov, V.D., Bychkova, L.V., Madej, J.}{2005}{A\&A}{430}{1143}

\refitem{
Donati J.-F., Morin J., Petit P., Delfosse X., Forveille T., Albert L.,
Auriere M., Cabanac R., Dintrans B., Fares R., Gastine T., Jardine M.M.,
Lignieres F., Paletou F., Ramirez Veles J.C., Theado S.} 
{2008}{MNRAS}{390}{545}

\refitem{Gershberg R.E.}{1972}{Ap\&SS}{19}{75}

\refitem{
Gershberg R.E., Katsova M.M., Lovkaya M.N., Terebizh A.V., Shakhovskaya N.I.}{1999}{aas}{139}{555}

\refitem{
Gershberg R.E., Terebizh A.V., Shlyapnikov A.A.}{2011}{BCrAO}{107}{11}

\refitem{
Hawley S.L., Fisher G.H., Simon T., Cully S.L., Deustua S.E., Jablonski M., Johns-Krull C.M.,
   Pettersen B.R., Smith V., Spiesman W.J., Valenti J.}{1995}{ApJ}{453}{464}

\refitem{
Lacy C.H., Moffet T.J., Evans D.S.}{1976}{ApJS}{30}{85}

\refitem{
Katsova M.M., Drake J.J., Livshits M.A.}{1999}{ApJ}{510}{986}

\refitem{
Katsova M.M. and Livshits M.A.}{2001}{Astronomical and Astrophysical Transactions}{20}{531}

\refitem{
Kowalski A.F., Hawley S.L., Holtzman J.A., Wisniewski J.P., Hilton E.J.}{2010}
   {ApJ}{714}{98}

\refitem{Krause F., Raedler K.-H.}{1980} 
   {Mean-field magnetohydrodynamics and dynamo theory, Oxford, Pergamon Press, Ltd.}{271}{}

\refitem{Moffett T.J.}{1974}{ApJS}{29}{1}

\refitem{
Morin J., Delfosse X., Donati J.-F., Dormy E., Forveille T., Jardine
   M.M., Petit P., Schrinner M.}{2011}{SF2A-2011}{}{503} 

\refitem{
Morin J., Dormy E., Schrinner M., Donati J.-F.}{2011}{MNRAS}{418}{L133}

\refitem{
Morin J., Donati J.-F., Petit P., Delfosse X., Forveille T., Albert L.,
Auriere M., Cabanac R., Dintrans B., Fares R., Gastine T., Jardine M.M.,
Lignieres F., Paletou F., Ramirez Velez J.C., TheadoS.}
{2008}{MNRAS}{390}{567} 

\refitem{
Morin J., Donati J.-F., Petit P., Delfosse X., Forveille T., Jardine M.M.}
{2010}{MNRAS}{407}{2269}

\refitem{ Pojma\'nski, G.}{1997}{Acta Astron.}{47}{467}

\refitem{Preston G.W.}{1971}{PASP}{83}{571}

\refitem{
Reid I.N., Hawley S.L.}{2005}{New Light on Dark Stars: Red Dwarfs}
{Low-Mass Stars, Brown Dwarfs}{(Chichester: Praxis Publishing Ltd.)}

\refitem{
Shibata K., Yokoyama T.}{1999}{ApJ}{526}{L49}

\refitem{
Shibata K., Yokoyama T.}{2002}{ApJ}{577}{422}

\refitem{
Stepanov A.V., Kopylova Yu.G., Tsap Yu.T., Kupriyanova E.G.}{2005}{Astronomy
Letters}{31}{612}

\refitem{Stibbs D.W.N.}{1950}{MNRAS}{110}{395}

\refitem{
Zhilyaev B. E., Tsap Yu. T., Andreev M. V., Stepanov A. V., Kopylova Yu. G.,
Gershberg R. E., Lovkaya M. N., Sergeev A. V., Verlyuk I. A., Stetsenko K. O.}
{2011}{Physics and kinematics of celestial bodies}{27}{75}


\end{references}
\end{document}